\documentclass[12pt,letterpaper]{article}

\pdfoutput=1
\usepackage{jheppub}
\usepackage{amsfonts}
\usepackage{amssymb}
\usepackage{graphicx}
\usepackage{latexsym}
\usepackage{amsmath}
\usepackage{float}
\usepackage{etoolbox}
\usepackage{color}

\newcommand\openone{\leavevmode\hbox{\small1\normalsize\kern-.33em1}}
\def\ee{\end{eqnarray}}

\newcommand{\di}{\partial}

\newcommand{\be}{\begin{eqnarray}}
\newcommand{\en}{\end{eqnarray}}
\newcommand{\bea}[1]{\left(\begin{array}{#1}}
\newcommand{\ena}{\end{array}\right)}
\newcommand{\ba}{\begin{eqnarray}}
\newcommand{\ea}{\end{eqnarray}}

\newcommand{\tr}{\text{Tr}}

\usepackage{hyperref}

\newcommand\lrpar{\raise .8ex\hbox{$^\leftrightarrow$} \hspace{-9pt}
\partial}
\newcommand\lpar{\raise .8ex\hbox{$^\leftarrow$} \hspace{-9pt}
\partial}
\newcommand\rpar{\raise .8ex\hbox{$^\rightarrow$} \hspace{-9pt}
\partial}

\makeatletter
\def\@fpheader{\relax}
\makeatother

\begin{document}

\subheader{\ }
\title{A solution of 2D QCD at Finite $N$ using a conformal basis}

\author[]{Emanuel Katz,}
\author[]{Gustavo Marques Tavares,}
\author[]{and Yiming Xu}

\affiliation[]{Physics Department, Boston University, Boston MA 02215}

\date{\today}

\abstract{We study 2D QCD with a fundamental fermion at small-$N$ using the recently proposed conformal basis approach.   We find that effective conformal dominance still holds, namely
that the spectrum converges efficiently, with high scaling-dimension operators decoupling exponentially quickly from the stable single-particle states.  Consequently,
for these stable bound states, accurate, analytic expressions for wavefunctions and parton distribution functions can be given, even for $N=3$.}


\maketitle

\section{Introduction}

In this paper we continue investigating the conformal basis technique for solving a gauge theory in 2D.   Previously, we have shown that the conformal basis is an efficient tool for 
finding the spectrum of lowest bound states in the context of 2D QCD with an adjoint fermion at large $N$ \cite{Katz:2013qua}.    In fact, basis states which correspond to high scaling-dimension operators
were found to decouple exponentially quickly from the low energy spectrum.  This phenomena, which we have named effective conformal dominance, is motivated
by holographic models of confinement \cite{Fitzpatrick:2013twa}.  The holographic analysis, however, assumed a large $N$ type theory, where bulk loops were neglected.  Thus, it was not clear whether 
this technique would be equally efficient for calculating the spectrum of a theory with a small number for colors.

Hence, we set out to explore 2D QCD with a single fundamental fermion for various values of colors ($N$=3, 6 and 1000).  After briefly reviewing the model, we present 
our results in Section \ref{results} where we discuss the single particle states, as well as the multi-particle states which include stable single-particle states and
arbitrary numbers of the decoupled massless scalar mode.  We then conclude in Section \ref{conclusion}.

\section{The Model}

The Lagrangian for the massless QCD in $1+1$ dimensions is
\be
\mathcal L = -\frac{1}{4}\tr (F^2) + i \bar \Psi \gamma^\mu D_\mu \Psi,
\label{eq:original lagrangian}
\ee
where $D_\mu \Psi = ( \partial_\mu - i g A_\mu^a T^a) \Psi$, with $T^a$ the generators of $SU(N)$ normalized such that $\tr (T^a T^b) = \delta^{ab}$. Using light-cone coordinates, $x^\pm = (x^0 \pm x^1)/\sqrt{2}$ and working in light-cone gauge $A_- = 0$ we can simplify the Lagrangian to
\be
\mathcal L = - \frac{1}{2}\tr\left( (\partial_- A_+)^2 \right) + i \psi^\dagger D_+ \psi + i \chi^\dagger \partial_- \chi.
\ee
The fields $\psi$ and $\chi$ are respectively the right and left movers, defined by $(1 \pm \gamma^3)\Psi$, where $\gamma^3 = i \gamma^0 \gamma^1$. The model has a chiral symmetry under which $\psi \rightarrow e^{-i \alpha} \psi$ and $\chi \rightarrow e^{ i \alpha} \chi$.

Choosing $x^+$ as our time coordinate we see that $A_-$ and $\chi$ are non-dynamical fields and can be integrated out. The momenta operators in terms of the dynamical field $\psi$ are given by
\be
\begin{split}
P^+ &= 2i \int dx^- \psi^\dagger \partial_- \psi \, , \\
P^- &=  -g^2 \int dx^-  \psi^\dag T^a \psi \frac{1}{\partial_-^{\, 2}} \, \psi^\dag T^a \psi.
\end{split}
\ee
Accordingly, the mass operator is given by
\be
M^2=2P^+P^-.
\ee
The spectrum of 2D QCD is solved through diagonalization of the mass matrix in a basis generated by conformal quasi-primary operators, truncated at dimension $\Delta_\text{max}$. These quasi-primary operators 
can be constructed from the fermion field $\psi$. The mass matrix elements are given by 
\be
\delta(p-q)M^2_{\Delta_1, \Delta_2}=\langle\mathcal{\tilde{O}}_{\Delta_1}(q)|2P^+P^-|\mathcal{\tilde{O}}_{\Delta_2}(p)\rangle,
\ee
with the Fourier transform of a quasi-primary operator defined as 
\be
\mathcal{\tilde{O}}_\Delta(p)=\int dx^{-} e^{ipx^{-}}\mathcal{O}_\Delta(x^{-}).
\ee
A detailed definition of the basis states is provided in the Appendix \ref{appprimary}.

In order to calculate the mass matrix we introduce the mode expansion at $x^+ = 0$,
\be
\psi_j(x^-) = \frac{1}{2 \sqrt{\pi}} \, \int_0^\infty \! dk^+ \left( e^{-i k^+ x^-} b_j (k^+) + e^{i k^+ x^-} a^\dag_j (k^+) \right),
\ee
where $j$ is the $SU(N)$ index in the fundamental representation. From the equal time anti-commutation relations for $\psi$,
\be
\{ \psi^\dagger_i(x^-), \psi_j(y^-) \} = \frac{1}{2} \delta_{ij} \delta(x^- - y^-),
\ee
one finds that the non-zero anti-commutators of the creation/annihilation operators are
\be
\begin{split}
\{ a_i(p^+) , a_j^\dag (k^+) \} = \delta_{ij} \delta(p^+ - k^+) \, , \\
\{ b_i(p^+) , b_j^\dag (k^+) \} = \delta_{ij} \delta(p^+ - k^+) \, .
\end{split}
\ee

The model is also invariant under charge conjugation, $\psi_i \rightarrow \psi^\dag_i$. This discrete symmetry breaks up the Hilbert space into two independent sectors which we denote by $C$-even and $C$-odd.

\section{Results}
\label{results}
In this section we present the results of the finite $N$ calculations.  As the Hamiltonian does not mix baryonic and mesonic operators, we will restrict ourselves to mesonic states in this paper.  We compare the spectra at different values of $N$. As expected we are able to identify both single-particle-states (the hadrons) and multi-particle states of hadrons in the spectra. In accordance with previous studies \cite{'tHooft:1974hx,Hornbostel:1988fb, Sugihara:1994xq, Berruto:2002gn}  we also identified a massless non-interacting state, $| B_0 \rangle$, even at finite $N$. 

In the large $N$ limit, the interaction between the single-particle states is $1/N$ suppressed \cite{'tHooft:1974hx, Callan:1975ps}. Therefore all mesons become stable in the infinite $N$ limit and the multi-particle states are identified with multiple non-interacting single-particle-states which are created by the multi-bilinear operators in the $\psi$ field (which we will refer to as multi-trace operators). Thus, the multi-particle states can be approximated by the eigenstates of a free effective Hamiltonian of various mass mesons, in a Hilbert space of non-interacting bosons truncated at the same operator dimension. At finite $N$ due to interactions between the bosons, 
we expect most mesons to acquire widths.  However, below the first multi-particle threshold, of two $|B_1\rangle$ mesons, all multi-particle states consist of a stable massive boson and multiple massless $|B_0 \rangle$ particles. There can be no interaction between the massless particle and any other states due to Coleman's theorem. We are thus able to match all states below the aforementioned threshold to states in the free effective Hamiltonian consisting of one massive hadron and multiple non-interacting massless $|B_0 \rangle$.

\subsection{Single Particle States}
\label{single-particle-states}

\begin{figure}
\begin{center}
\includegraphics[width=0.7\textwidth]{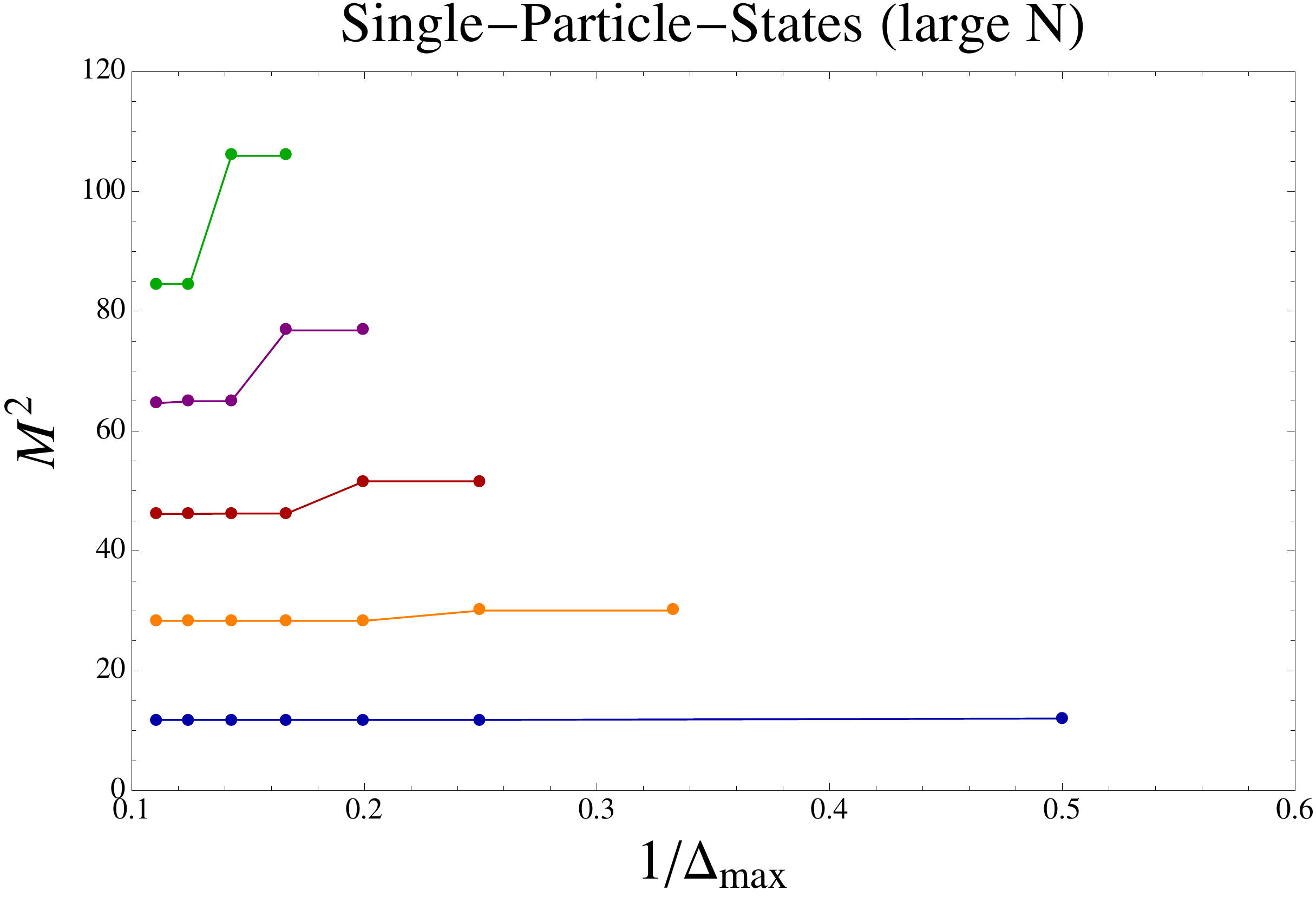}
\includegraphics[width=0.7\textwidth]{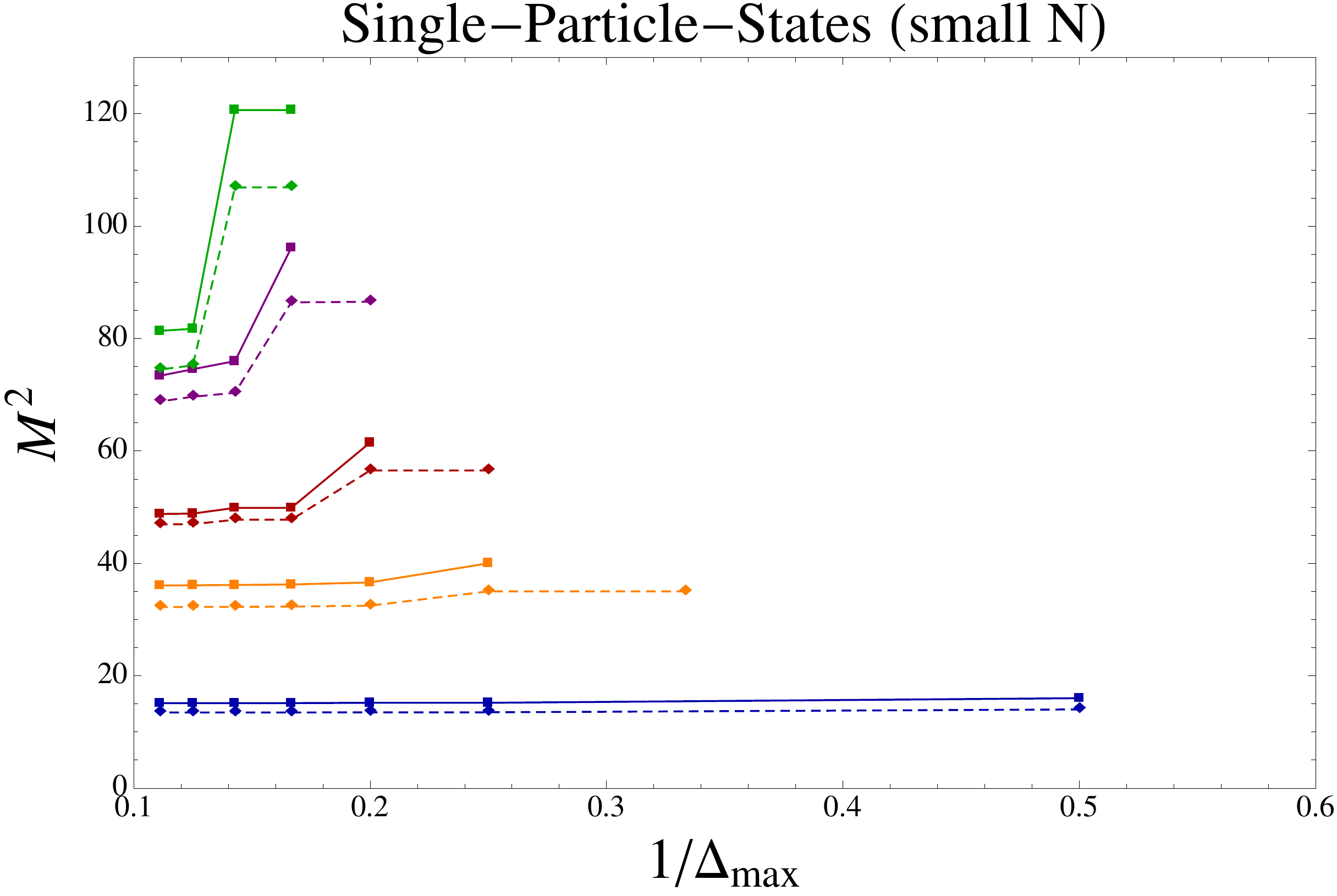}
\end{center}
\caption{The spectra of the lowest mesonic states at large $N$ $(N=1000)$ and small $N$ $(N=3 \ \text{solid lines}, \ N= 6 \ \text{dashed lines})$, calculated up to $\Delta_\text{max}=9$. The unit of $M^2$ is $g^2/2\pi$. The lowest three states in all cases are decoupled from the high dimensional operators. The "Regge trajectory" of the spectrum at large $N$ can be seen in the first plot.}
\label{fig:singleparticle-spectrum}
\end{figure}

We begin by investigating the single-particle states. In the large $N$ limit these hadronic states are the ones obtained by solving the 't Hooft equation, which can be obtained in our approach by neglecting sub-leading $N$ correction and using only the single trace operators to generate the basis. In order to identify the single particle states we plot their mass square as a function of $\Delta_\text{max}$, the dimension of the highest dimension operators included in our basis. The single particle states are identified as the ones for which the mass becomes constant at higher $\Delta_\text{max}$. In Figure~\ref{fig:singleparticle-spectrum} we plot the masses of all single particle states we identified as a function of $\Delta_\text{max}$ for $N =  3, \, 6$ and $1000$. The mass square is in units of $g^2/2\pi$. At large $N\ (N=1000)$ our result matches with that of \cite{'tHooft:1974hx}. One can see from the figure that all stable single particle states obtained at finite $N$ smoothly match to a corresponding large $N$ state. Our result for the mass of the lightest meson for $N=3$ is in good agreement (within 10\% of) earlier numerical results obtained using discrete light-cone quantization~\cite{Hornbostel:1988fb}.

\begin{figure}
\begin{center}
\includegraphics[width=1.\textwidth]{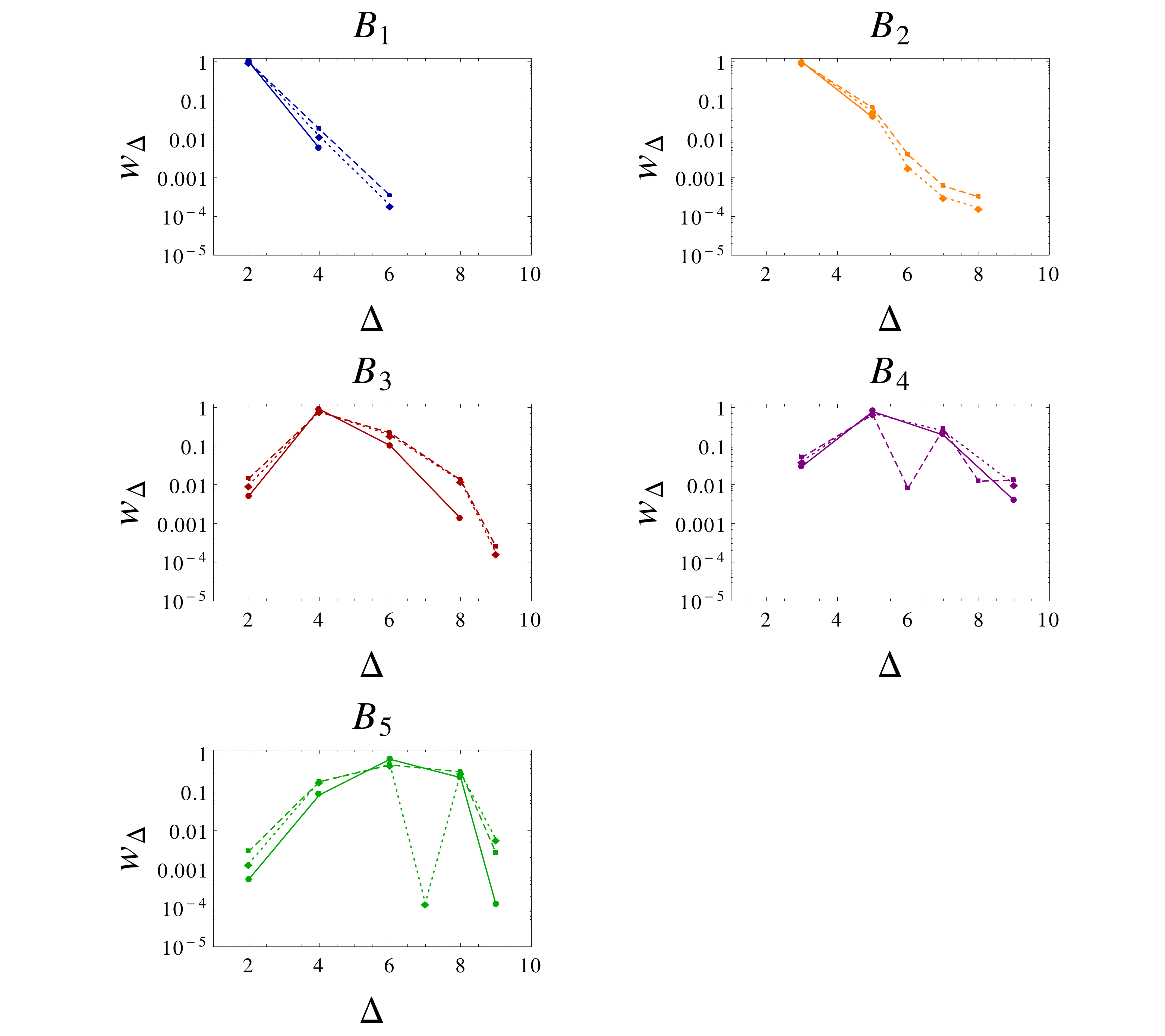}
\end{center}
\caption{Plots of the weight of states as a function of operator dimension $\Delta$, showing the decoupling of the lowest five single-particle-states. The circles connected by solid lines correspond to the $N=1000$ case, the diamond points - dotted lines to $N=6$ and the square points - dashed lines to $N=3$. We did not include points for which the weight is smaller than $10^{-4}$.}
\label{fig:singleparticle-convergence}
\end{figure}

In order to test if the single particle states satisfy the effective conformal dominance hypothesis, we calculate, for each state, the contribution from operators of a given dimension $\Delta$. For example, for a state given by the general expression
\be
| \psi \rangle = \sum_{\Delta} \sum_{i} c_{\Delta, i}\mathcal{O}_{\Delta, \, i} | \Omega \rangle,
\ee
where $i$ labels the different independent operators of dimension $\Delta$, we define the weight of this state at dimension $\Delta$ as
\be
w_\Delta = \sum_{i} |c_{\Delta, i}|^2.
\ee
In Figure~\ref{fig:singleparticle-convergence} we plot the weight at each dimension for all the single particle states identified in our spectrum. We see that, at $\Delta_\text{max}=9$, the lowest three single-particle-states decouple exponentially fast from the high dimensional operators, as expected from the effective conformal dominance hypothesis. Another interesting aspect of Figure~\ref{fig:singleparticle-convergence}  is that the slope associated with this decoupling varies slowly with $N$. These three states have masses below the threshold of $4M^2_{B_1}$, so they are stable. The mass of the next state, $M^2_{B_4}$, lies in the regime between $4M^2_{B_1}$ and $\left(M_{B_1}+M_{B_2}\right)^2$. Nevertheless, because it is in the $C$-odd sector,  symmetry forbids its decay into two $|B_1\rangle$ particles and thus it is also stable. $|B_5\rangle$ is the first unstable single-particle-state that can decay to $|B_1\rangle\otimes |B_1\rangle$. However, given that $M_{B_5}$ is larger than $M_{B_4}$, and even the stable state $|B_4\rangle$ is not completely decoupled from higher dimensional operators, it is hard to explore the unstable state using the current truncated basis.  It would be interesting to investigate the appearance of a width with a larger basis in the future.

Because of effective conformal dominance at finite $N$, the wave-functions of the single-particles that have decoupled can be well approximated by the lowest operators. For instance, at $N=3$, the lightest state $|B_1\rangle$ is at $98\%$ created by the lowest two operators in the $C$-even sector. Therefore, to this precision, the lightest meson is given by a simple expression, which can be obtained from an analytic calculation,
\be
|B_1\rangle=0.81\left(\sqrt{3}(\di\psi^\dag\psi-\psi^\dag\di\psi)\right)|\Omega\rangle-0.57\left(\frac{3}{\sqrt{2}}(\psi^\dag\psi)^2\right)|\Omega\rangle.
\ee
Here the numbers in the parentheses account for the normalization of the operators. The above expression contains all the information about the state $|B_1\rangle$. From it one can, for example, easily obtain the probability of finding a quark with momentum fraction $x$
\be
P(x)=1.96 \left(2 x -1\right)^2+1.95 \left(\frac{1}{2}-x+\frac{x^2}{2}\right).
\ee
At large $N$ this state is dominated by the stress-tensor operator, $\di\psi^\dag\psi-\psi^\dag\di\psi$, and it becomes the first massive mesonic state in the large $N$ 't Hooft model.   At finite $N$ however, we
see that the current squared operator $(\psi^\dag\psi)^2$ makes comparable contribution.

\subsection{Massless Sector}

\begin{table}[]
\centering
\begin{tabular}{|c|c|c|c|c|c|c|c|c|c|}\hline $\Delta_\text{max}$ & 1&2& 3 & 4 & 5 & 6 & 7 & 8 & 9 \\\hline\hline Masslesss C- even states $(N=1000)$ & 0&1& 1 & 3 & 3 & 6 & 7 & 12 & 14 \\\hline Masslesss C- even states $(N=3)$ & 0&1& 1 & 3 & 3 & 6 & 7 & 12 & 14 \\\hline $\mathbb Z_2$ even bosonic operators &0&1&1 & 3 & 3 & 6 & 7 & 12 & 14 \\\hline\hline Masslesss  C- odd states $(N=1000)$ & 1&1&  2 & 2 & 4 & 5 & 8 & 10 & 16 \\\hline Masslesss C- odd states $(N=3)$ & 1&1&  2 & 2 & 4 & 5 & 8 & 10 & 16 \\\hline $\mathbb Z_2$ odd bosonic operators &  1&1& 2 & 2 & 4 & 5 & 8 & 10 & 16 
\\\hline \end{tabular}
\caption{Numbers of massless states in the even and odd sectors of C-symmetry and the numbers of bosonic quasi-primary operators with corresponding $\mathbb Z_2$ symmetry, at each operator dimension.} \label{table:masslesstevennum}
\end{table}

The spectrum of this model, at arbitrary $N$, includes a sector of massless non-interacting states, as had been previously discovered in the limit of vanishing quark mass by numerical methods \cite{Hornbostel:1988fb, Sugihara:1994xq, Berruto:2002gn}. In particular one can show that the state created by the operator $\psi^\dagger \psi$ is an eigenstate of $P^-$ with zero eigenvalue and therefore corresponds to a non-interacting massless state. Because all our states are right movers, any multi-particle state made of an arbitrary number of non-interacting massless right movers will also be massless.

This massless, non-interacting sector can be completely described by a theory of a single free scalar field with a shift symmetry. The charge conjugation symmetry of the original model is mapped to a $\mathbb Z_2$ symmetry of the scalar field, $\phi \rightarrow - \phi$. In order to match the massless sector of the original theory to this free CFT we compare the number of quasi-primary operators in the free theory
of dimension less than $\Delta_\text{max}$ with the number of massless states obtained in the original model using the basis of  fermionic operators up to the same dimension cutoff $\Delta_\text{max}$.

In Table \ref{table:masslesstevennum}  we list the number of massless states in the original theory as a function of the dimension cutoff $\Delta_\text{max}$, for different values of $N$, and for the $C$-even and odd sectors. In the same table we also show the number of primary operators in the free scalar CFT with dimensions less than $\Delta_\text{max}$, grouped by their charge under the $\mathbb Z_2$ symmetry. For example, at $\Delta_\text{max}=4$, the three $\mathbb Z_2$-even bosonic operators are $(\di\phi)^2$, $(\di\phi)P^{(1,1)}_{2}\! \left(  \overleftarrow \partial - \overrightarrow \partial \right)(\di\phi)$ and $(\di\phi)^4$, with $P^{(1,1)}_{2}$ a Jacobi polynomial. The number of massless states in the original theory is independent of $N$. It exactly matches the number predicted by the free CFT at any given dimension. This is a non-trivial check that the massless sector is completely described in terms of a free CFT of a single scalar field with the identification $\psi^\dagger \psi \rightarrow \partial \phi$. It shows that there is only one non-interacting massless single-particle $|B_0\rangle$ in the spectrum, all other massless states are states with multiple $|B_0\rangle$ particles.

\subsection{Multi-particle States}
\label{multi-part}

\begin{figure}
\begin{center}
\includegraphics[width=0.65\textwidth]{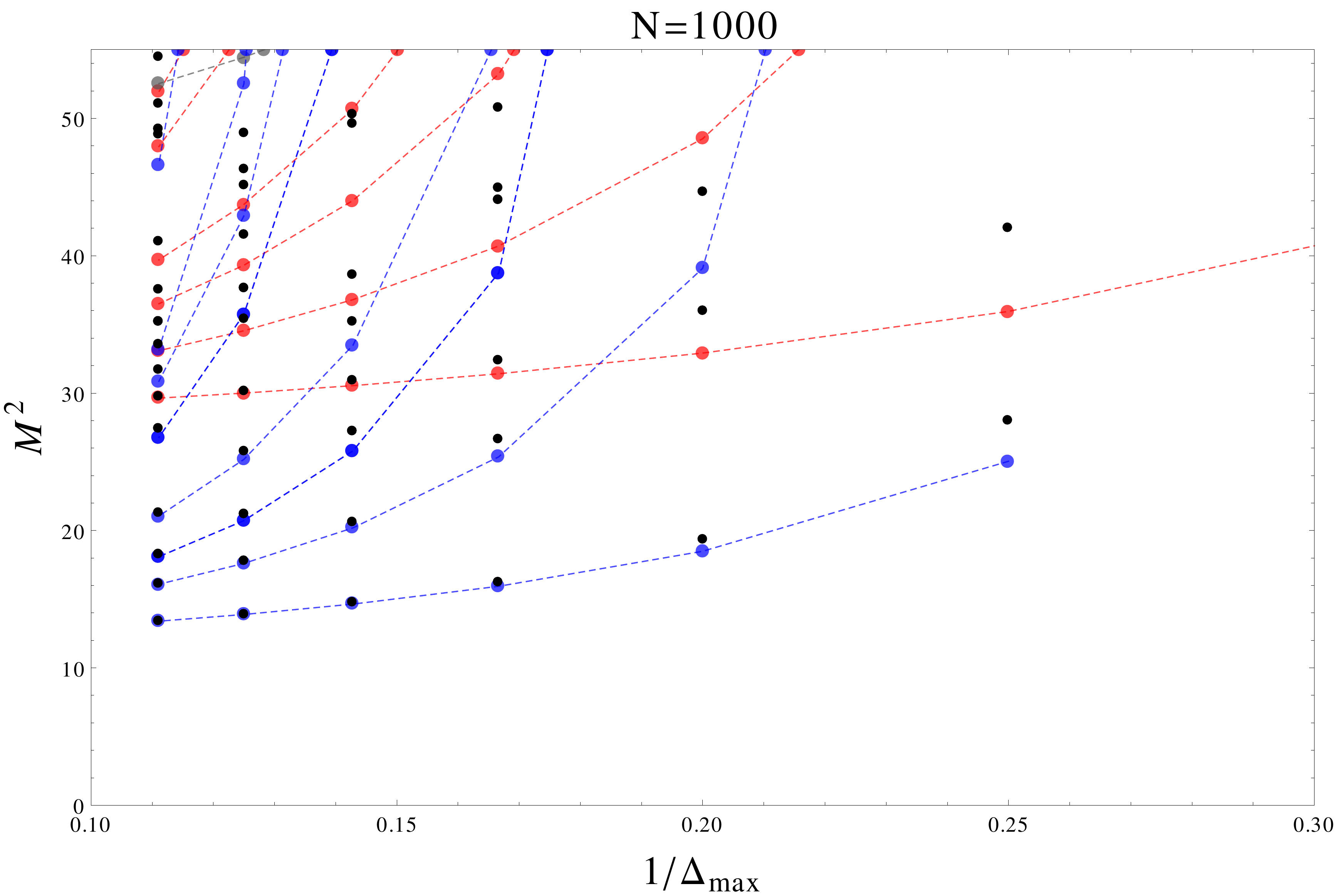}
\includegraphics[width=0.65\textwidth]{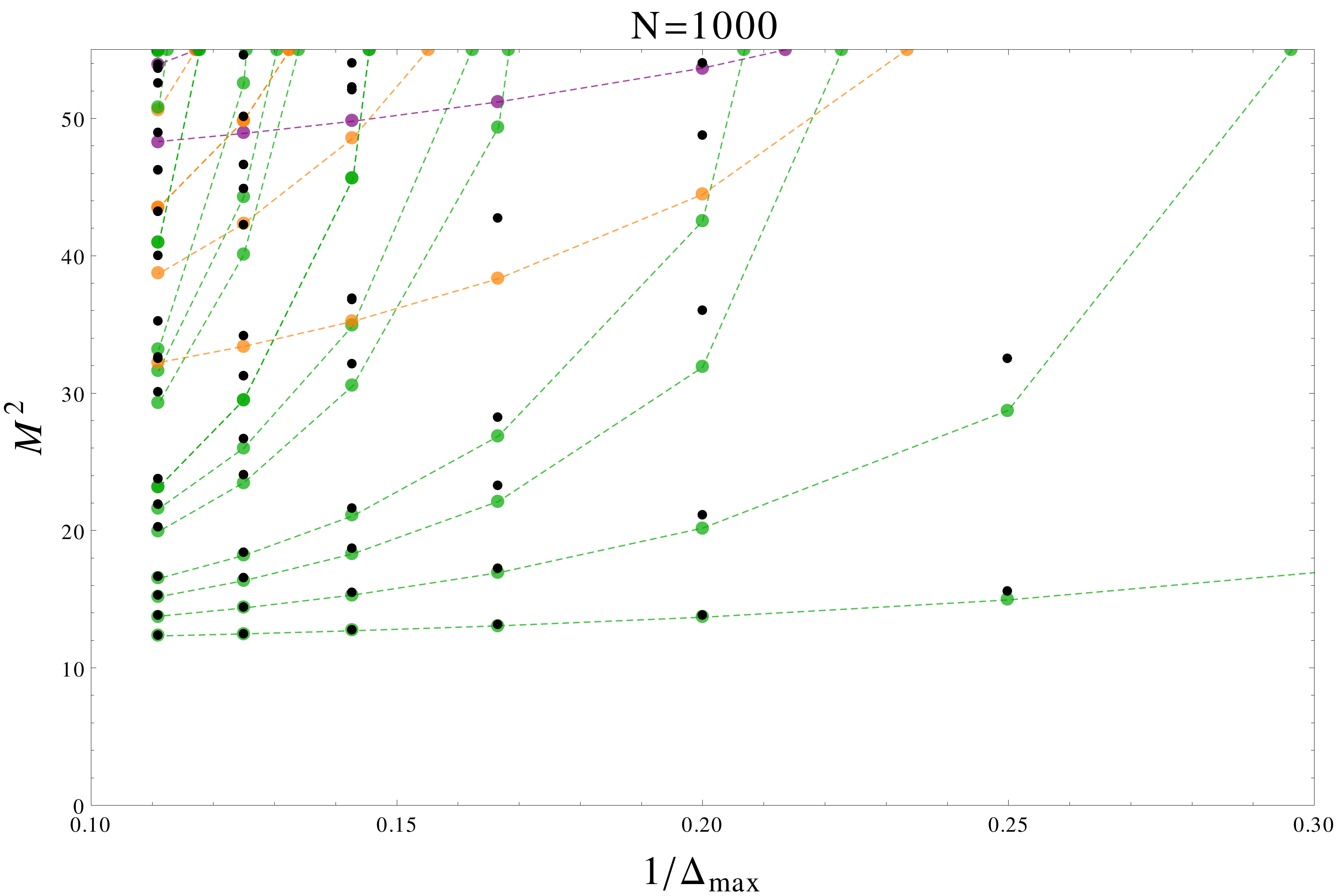}
\end{center}
\caption{Matching of the spectrum of multi-particle-states, $|B_1\rangle\otimes n_1 |B_0\rangle$, $|B_2\rangle\otimes n_2 |B_0\rangle$ and $|B_3\rangle\otimes n_3 |B_0\rangle$ ($n_i\geqslant 1$, $n_i\in \mathbb Z$), with the non-interacting multi-meson spectrum, at $N=1000$. States are shown below the threshold of $4M^2_{B_1}\sim 45$. The black dots show the spectrum of the finite $N$ 't Hooft model, which is obtained from a fermonic operator basis, whereas the colored dots indicate the mass eigenvalues obtained from diagonalizing a free Hamiltonian using a bosonic basis. The charge-conjugate $C$-even sector is shown in the first plot (blue: $|B_1\rangle \otimes n_1 |B_0\rangle$, red: $|B_2\rangle \otimes n_2 |B_0\rangle$, gray: $|B_3\rangle\otimes n_3 |B_0\rangle$).  The second plot is for the $C$-odd sector (green: $|B_1\rangle\otimes n_1 |B_0\rangle$, orange: $|B_2\rangle\otimes n_2 |B_0\rangle$, purple: $|B_3\rangle\otimes n_3 |B_0\rangle$).}
\label{fig:multiparticle-convergence-N1000}
\end{figure}

\begin{figure}
\begin{center}
\includegraphics[width=0.65\textwidth]{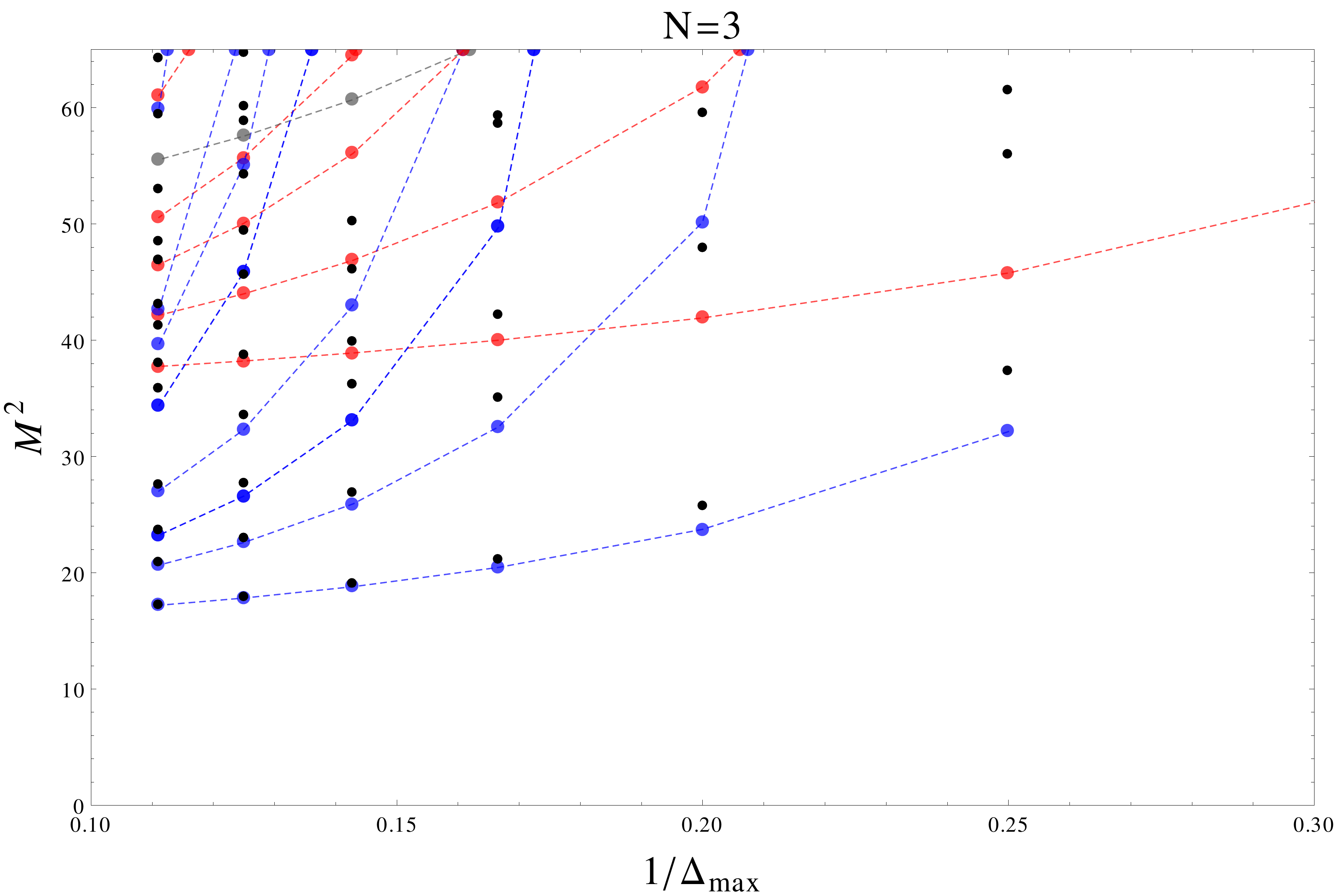}
\includegraphics[width=0.65\textwidth]{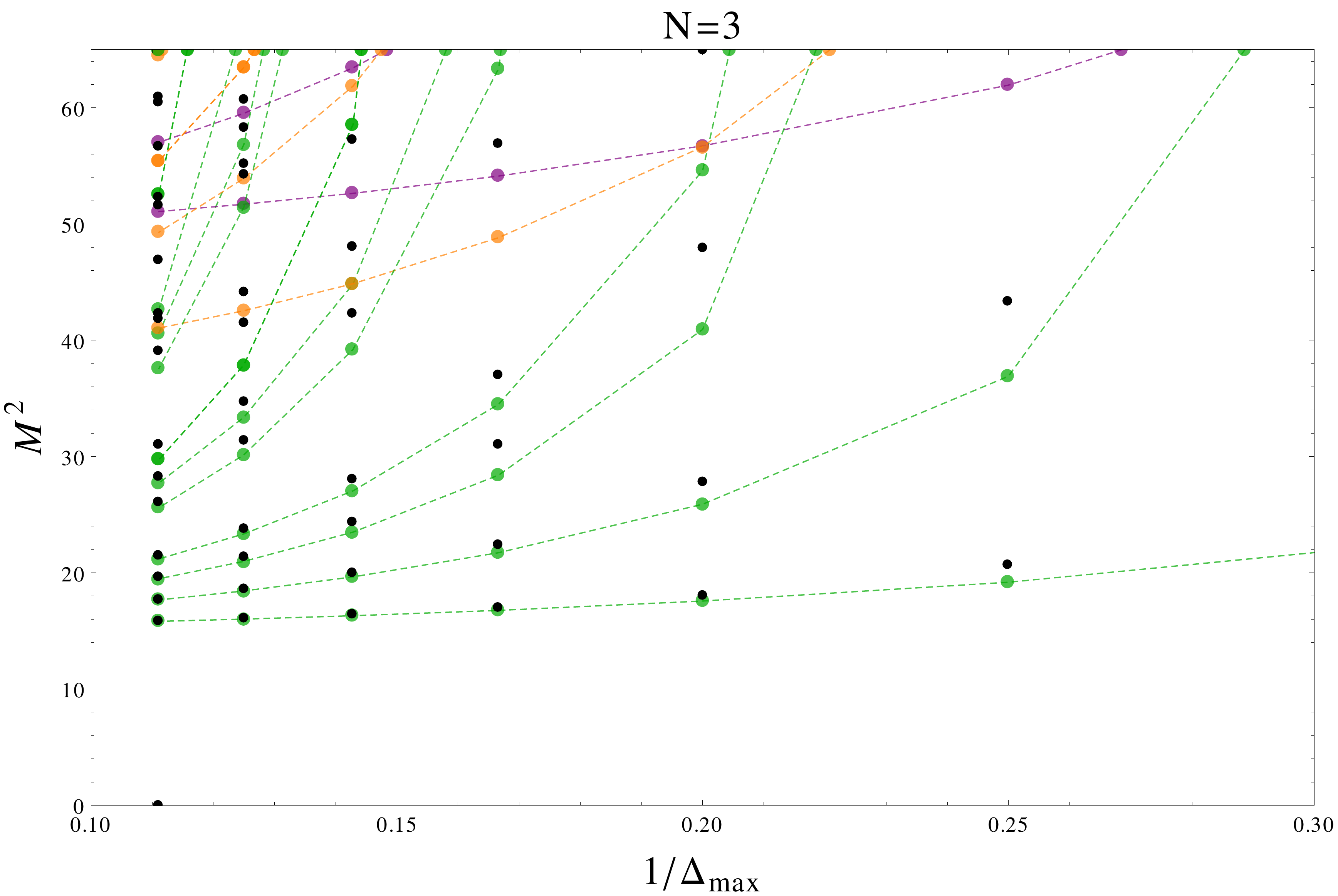}
\end{center}
\caption{Matching of the spectrum of multi-particle-states, $|B_1\rangle\otimes n_1 |B_0\rangle$, $|B_2\rangle \otimes n_2 |B_0\rangle$ and $|B_3\rangle \otimes n_3 |B_0\rangle$ ($n_i\geqslant 1$, $n_i\in \mathbb Z$), with the non-interacting multi-meson spectrum, at N=3. States are shown below the threshold of $4M^2_{B_1}\sim 60$. The color code is similar to that of the case of $N=1000$ (Figure \ref{fig:multiparticle-convergence-N1000}). }
\label{fig:multiparticle-convergence-N3}
\end{figure}

The effective conformal dominance hypothesis implies that, at certain cutoff dimension $\Delta_\text{max}$, the truncated approximation to all the states in the spectrum below a certain mass scale will converge exponentially. This is the case for the states identified as single-particles. However, below the highest mass, $M_{\Delta_\text{max}}$, of the single-particle-states that have already converged, we find additional mass eigenstates whose masses do not asymptote to constant values as a function of the cutoff $\Delta_\text{max}$. These states are identified as multi-particle states and can be viewed as the attempt of the truncated basis to reproduce the full continuum of the Hilbert space of gauge-singlet states. 

The continuum spectrum represents states with multiple mesons, i.e. several of the single-particle states studied in the earlier section. Similar to the case of the massless sector, these multi-particle states can be described, at least in the limit that their interaction is sufficiently weak, by an effective model of bosonic degrees of freedom. For each meson in the single-particle state there is a bosonic field associated to it in the model. By using the conformal basis approach to calculate the spectrum in this effective bosonic description we expect that, as $\Delta_\text{max}$ increases the truncated spectrum in the bosonic description will match the one in the full fermionic theory for states with masses lower than $M_{\Delta_\text{max}}$, the mass of the heaviest meson that has converged. The construction of the bosonic basis is described in the Appendix~\ref{appprimary}.

As expected, the discreteness of our truncated basis can only approximate the continuum spectrum, even in the simpler case of a free theory. In Appendix~\ref{app density} we explore how one recovers the appropriate continuum 
density of states with the conformal basis approach in the case of free bosons.  The continuum density of states is well approximated once $\Delta_\text{max}$ becomes large.  However, at the values of $\Delta_\text{max}$
we are working with in this paper, in order to understand the multi-particle states in the truncated basis, it is more useful to compare the truncated 2D QCD spectrum to the corresponding truncated free boson model. 
The matching of the full spectrum to the expected free boson model is evidence that we have correctly reproduced the continuum physics.

At large $N$, the interaction between the mesons is of order $1/N$, and the spectrum is that of a free Hamiltonian of multiple bosons. The mass matrix is given by
\be
M^2=2P^+P^{-}=\frac{M^2_1}{x_1}+\frac{M^2_2}{x_2}+...+\frac{M^2_n}{x_n}
\label{freeHamiltonian}\ee
where $M_i$ are the masses of the mesons, which have been calculated in Section~\ref{single-particle-states}, and the parton variables are defined as $x_i=P^{-}_i/P^{-}$ subjected to the constraint $x_1+x_2+...+x_n=1$.

At finite $N$ the free Hamiltonian eq. (\ref{freeHamiltonian}) is distorted by particles acquiring widths. One therefore expects that it will not correctly model the spectrum. There is an exception, however, for states containing the 
massless state $|B_0\rangle$, which has no interactions with other massive states. The free Hamiltonian is thus exact for multi-particle states with one stable meson and any number of  massless $|B_0\rangle$ particles.

In Figures~\ref{fig:multiparticle-convergence-N1000} and  \ref{fig:multiparticle-convergence-N3} we show that, below the threshold of $M^2=4M^2_{B_1}$ in the QCD model spectrum, all the eigenstates besides the single-particle-states can be matched with the excitations $|B_1\rangle\otimes n_1 |B_0\rangle$, $|B_2\rangle\otimes n_2 |B_0\rangle$ and $|B_3\rangle\otimes n_3 |B_0\rangle$ ($n_i\geqslant 1$, $n_i\in \mathbb Z$). Because $|B_1\rangle$ and $|B_3\rangle$ are in the $C$-even sector, whereas $|B_0\rangle$ and $|B_2\rangle$ are in the odd sector, multi-particle-states in the $C$-even sector should be those where $n_1$ and $n_3$ are odd integers and $n_2$ is even. The opposite applies to multi-particle-states in the $C$-odd sector. Here we demonstrate the matching in the cases of $N=1000$ and $N=3$. The black dots are the 2D QCD model mass eigenvalues calculated at a given $\Delta_\text{max}$. The colored dots, connected by the dashed lines which indicate their trend, show the spectrum of the free Hamiltonian ~(\ref{freeHamiltonian}), using the bosonic basis truncated at the same $\Delta_\text{max}$, with the right $C$-symmetry. At the highest $\Delta_\text{max}=9$ we have reached in our calculation there is a one-to-one correspondence between the black and colored dots. In fact for the low-lying states it is easy to see that the variation of mass as a function of $\Delta_{\max}$ of the original state follows that of the non-interacting spectrum. This indicates again the matching of the two spectra. Hence we can identify all the eigenstates of the QCD model calculated using the truncated fermionic basis.

One can also see from Figure~\ref{fig:multiparticle-convergence-N1000} that at $M^2$ around $45$, which is the mass square of  the heaviest single-particle-state that has converged at $\Delta_\text{max}=9$, the black dots start to be significantly higher than the corresponding colored dots representing the free Hamiltonian description of the same state. This suggests that at these masses our full theory calculation errors have become large and the corresponding states have not yet converged. The threshold at which we expected to observe states with two $|B_1\rangle$ is larger than $M^2 = 45$, and likely beyond the regime of convergence at our current $\Delta_\text{max}$.

\section{Conclusion}
\label{conclusion}
In this paper we showed that the effective conformal dominance hypothesis continues to be true in 2D QCD at finite $N$. As expected, at finite $N$, the decoupling with respect to the cutoff dimension is slower than in the case of infinite $N$ , but qualitatively we do not observe a significant change in the exponential suppression of the high dimensional operator contribution to a light state. The conformal basis approach remains reliable in computing the low energy spectrum. Both the single particle spectrum and the continuum in the model are identified with accuracy characterized by $e^{-\Delta_{\text{max}}}$.

Some properties of the conformal basis approach can be summarized as follows. First, it provides a way to non-perturbatively define the 2D gauge theory. Second, the basis is a discretization that naturally uses CFT discreteness without the need to introduce additional ``external'' deformations of the theory, which is different from lattice gauge theories or DLCQ methods. Third, it is an effective method for computing the low energy spectrum, where the light states can be understood analytically.  

The method has some resemblance to the so-called Truncated Conformal Space Approach (TCSA), which has been applied to certain 2D theories on a cylinder
of fixed radius, with a cutoff imposed on the maximum energy of excitations \cite{Yurov:1989yu}.  
The KK-modes of the circle in this context are related to conformal states through radial quantization.  TCSA has been traditionally applied to CFT deformations by a local relevant operator.   
The basis used at a given energy cutoff (the equivalent of our $\Delta_\text{max}$) is effectively larger as it includes both quasi-primaries and their global conformal descendants 
(as $P_\mu$ translation symmetry is broken by the circle).  
The continuum spectrum is then obtained as the radius is taken to be large (in this sense it is similar to DLCQ).
In cases where the vacuum energy requires renormalization, the TCSA method faces a challenge, as currently there is no simple, systematic, renormalization procedure (see \cite{Giokas:2011ix} for discussion). 
In the 2D QCD case studied here, the deformation of the free fermion CFT is technically non-local (once the gluon has been integrated out).  As our method preserves translation along $x^-$ explicitly, we did not need to include basis states that correspond to the descendants of the quasi-primaries.
In addition, on a spatial circle, quantization would likely involve vacuum renormalization, a complication avoided in our light-cone framework.  Thus, there does not seem to be a straight forward way of relating our method to TCSA, although it would be interesting to investigate the 
connection between the two in the future.

Generalizing this method to higher dimensional theories and to cases with dimensional transmutation would be very interesting.

\section*{Acknowledgments}
We thank Claudio Rebbi and Slava Rychkov for useful discussions. The work of EK and YX was supported in part by DOE grant DEFG02-01ER-40676. GMT acknowledges support from a DOE High Energy Physics Fellowship.

\appendix
\label{appendix}
\section{The Quasi-primary Operator Basis}
\label{appprimary}
In this appendix we explain briefly the construction of the single-traced and multi-traced quasi-primary operators of the 2D QCD model with a fundamental fermion. These are composite operators of the fermionic field. We also construct the quasi-primary operators of the bosonic fields which are used in the description of the continuum spectrum in terms of multiple free massive particles as described in the text.
These primary operators are obtained from solving the Killing equations. A similar and detailed description of the solutions can be found in \cite{Katz:2013qua}.

We are interested in quasi-primary operator with $2k$ fermions, at dimension $n+k/2$, that can be written as
\be
\mathcal{O}(x^-)=\frac{1}{N^k}\sum_{\sum {s_i}=n} c_{s_1, s_2, ..., s_{2k}}\left(\di^{s_1}\psi_{i_1}^\dag\di^{s_2}\psi_{i_1}\right)...\left(\di^{s_{2k-1}}\psi_{i_k}^\dag\di^{s_{2k}}\psi_{i_k}\right),
\label{primaryop}
\ee
with $i$'s the $SU(N)$ color indices. Since in the conformal limit, $\partial_+ \psi = 0$, by the equations of motion, one needs to consider only derivatives with respect to the ``space-like'' coordinate $x^{-}. $\footnote{Hereafter, we will drop the superscript $-$ when there is no ambiguity.} Furthermore, in the massless limit the right moving state $\psi$ decouples from the left moving one $\chi$, therefore to calculate the spectrum one needs to consider only composite operators of the $\psi$ field.

For testing the decoupling of operators at finite $N$, we restrict our basis to the mesonic states that contain color-contracted pairs of $\psi^\dag\psi$ fields. The operators given in eq.~(\ref{primaryop}) form a complete basis of the primary operators that are neutral under the chiral transformation  $\psi\rightarrow e^{i\alpha}\psi$. Any chiral-charged operator of the form $$\epsilon^{i_1, i_2, ... i_N}\di^{s_1}\psi_{i_1}\di^{s_2}\psi_{i_2}...\di^{s_N}\psi_{i_N}\left(\di^{r_1}\psi_{j_1}^\dag\di^{r_2}\psi_{j_1}\right)...\left(\di^{r_{2l-1}}\psi_{j_l}^\dag\di^{r_{2l}}\psi_{j_l}\right)$$ cannot interfere through the Hamitonian with the mesonic states because they have different charges under chiral symmetry. In addition, one can prove that any chiral-neutral ``baryon'' operator $$\epsilon^{i_1, i_2, ... i_N}\di^{s_1}\psi^\dag_{i_1}\di^{s_2}\psi^\dag_{i_2}...\di^{s_N}\psi^\dag_{i_N}\epsilon^{j_1, j_2, ... j_N}\di^{r_1}\psi_{j_1}\di^{r_2}\psi_{j_2}...\di^{r_N}\psi_{j_N},$$ can be written as a combination of meson operators and therefore is already included in eq.~(\ref{primaryop}).

The coefficients $ c_{s_1, s_2, ..., s_{2n}}$ are solved by imposing the Killing equation
\be
[K_-, \mathcal{O}_{n+k/2}(x^-)]=i\left((x^-)^2\partial_-+x^-(2n+k)\right)\mathcal{O}_{n+k/2}(x^-). 
\ee
A generic solution to the equation is given in \cite{Katz:2013qua}.

The 2D QCD Hamiltonian has charge conjugation symmetry: $\psi^\dag\leftrightarrow \psi$. This divides the Hibert space generated by the operators into an even and an odd sector under the symmetry. Another symmetry of the quasi-primary operator is the permutation of the pairs of the color contracted fermion-anti-fermion fields $\di^{s_{2i-1}}\psi^\dag\di^{s_{2i}}\psi$. This implies that the coefficient $c_{s_1, s_2, ..., s_{2n}}$ is invariant under the exchange of indices $s_{2i-1}\leftrightarrow s_{2j-1}$ and $s_{2i}\leftrightarrow s_{2j}$. The two symmetries impose further constraint on the coefficient $c_{s_1, s_2, ..., s_{2n}}$.

An operator of dimension $\Delta$ is normalized such that
\be
\lim_{x^{-}\rightarrow \infty} \left(x^{-}\right)^{2\Delta} \langle \mathcal{O}_\Delta(x^{-})\mathcal{O}_\Delta(0)\rangle =(-1)^\Delta \Gamma\left(2\Delta-1\right).
\ee
The factor $(-1)^\Delta$ on the right hand side is coming from having the operators on the same time slice of $x^+$. 
A state associated with $\mathcal{O}_\Delta$  is defined through the Fourier transform  $\mathcal{\tilde{O}}_\Delta(p)=\int dx^{-} e^{ipx^{-}}\mathcal{O}_\Delta(x^{-})$,
\be
|\mathcal{\tilde{O}}_\Delta(p)\rangle=\frac{1}{p^{\Delta-1/2}}\mathcal{\tilde{O}}_\Delta(p)|0\rangle.
\ee
The state, labeled by the Casimir and momenta, is thus normalized as
\be
\langle\mathcal{\tilde{O}}_\Delta(q)|\mathcal{\tilde{O}}_\Delta(p)\rangle=\delta(p-q).
\label{innerprod}
\ee

To evaluate the inner product eq.~(\ref{innerprod}), given that the quasi-primary operators are composite operators, we insert the identity with a complete set of momentum eigenstates, i.e., 
\be
\mathbf{1}=\sum_{k}\int \prod^k_{i=1}\ d p_i d p^{\prime}_i |p_1, p^{\prime}_1,..., p_k, p^{\prime}_k\rangle \langle \widetilde{p_1, p^{\prime}_1,..., p_k, p^{\prime}_k}|.
\ee
Here the momentum eigenstate kets are defined by
\be
|p_1, p^\prime_1,..., p_k, p^\prime_k\rangle = \frac{1}{N^k} a^\dag_{i_1}(p_1) b^\dag_{i_1}(p_1^\prime) ... a^\dag_{i_k}(p_k) b^\dag_{i_k}(p_k^\prime) | 0 \rangle,
\ee
with $a^\dag$ and $b^\dag$ the creation operators for quarks and anti-quarks and where repeated color indices $i_n$ are contracted.  The bra $\langle \widetilde{p_1, p^\prime_1,..., p_k, p^\prime_k}|$ is not the Hermitian conjugate of $|p_1, p^\prime_1,..., p_k, p^\prime_k\rangle$ as usual but it contains also all the subleading terms in $1/N$, such that for any momentum eigenstate $|q_1, q^\prime_1, ..., q_N, q^\prime_N\rangle$, 
\be
\begin{split}
&\sum_{k}\int \prod^k_{i=1}\ d p_i d p^\prime_i |p_1, p^\prime_1,..., p_k, p^\prime_k\rangle \langle \widetilde{p_1, p^\prime_1,..., p_k, p^\prime_k}|q_1, q^\prime_1, ..., q_N, q^\prime_N\rangle\\
&=|q_1, q^\prime_1, ..., q_N, q^\prime_N\rangle.
\end{split}
\ee
As an example, for a state with two pairs of quark and anti-quark, $\langle \widetilde{p_1, p^\prime_1, p_2, p^\prime_2}|=\frac{N^2}{2(N^2-1)}\langle p_1, p^\prime_1, p_2, p^\prime_2|+\frac{N}{2(N^2-1)}\langle p_2, p^\prime_1, p_1, p^\prime_2|$.

Consequently
\be
\begin{split}
&\langle\mathcal{\tilde{O}}_\Delta(q)|\mathcal{\tilde{O}}_\Delta(p)\rangle\\
&=\sum_{k}\int \prod^k_{i=1}\ d p_i d p^\prime_i\langle\mathcal{\tilde{O}}_\Delta(q)|p_1, p^\prime_1,..., p_k, p^\prime_k\rangle \langle \widetilde{p_1, p^\prime_1,..., p_k, p^\prime_k}|\mathcal{\tilde{O}}_\Delta(p)\rangle\\
&\equiv \int \prod^N_{i=1}\ d p_i d p^\prime_i\delta(p-\sum p_i)\delta(q-\sum p_i)  \tilde{f}(p_1, p^\prime_1, ..., p_N, p^\prime_N)f^*(p_1, p^\prime_1, ..., p_N, p^\prime_N).
\end{split}
\ee
 The function $f(p_1, p^\prime_1, ..., p_N, p^\prime_N)$, generically a polynomial of the momenta $p_i$ and $p^\prime_i$, is defined for the operator $\mathcal{O}_\Delta$ as $f(p_1, p^\prime_1, ..., p_N, p^\prime_N)\equiv \langle p_1, p^\prime_1, ..., p_N, p^\prime_N|\mathcal{\tilde{O}}_\Delta(p)\rangle$. Similar definition applies to $\tilde{f}$. The normalization of a state, eq.~(\ref{innerprod}) , thus translates into the condition of the polynomial functions, that
\be
 \int \prod^N_{i=1}\ d p_i d p^\prime_i\tilde{f}(p_1, p^\prime_1, ..., p_N, p^\prime_N)f^*(p_1, p^\prime_1, ..., p_N, p^\prime_N)=1.
  \ee
Here the integral is on the simplex $\sum p_i+\sum p^\prime_i=p$. 

The quasi-primaries with different operator dimension are orthogonal. We need to orthogonalize only the operators with the same dimension, by a Gram-Schmidt procedure, to obtain an orthonormal basis.

The quasi-primary operators of lowest dimension in our basis are
\be
\begin{split}
\mathcal{O}^{(1)}&\sim\psi^\dag\psi,\\
\mathcal{O}^{(2)}&\sim(\di\psi^\dag)\psi-\psi^\dag\di\psi,\\
\mathcal{O}^{(3)}&\sim(\psi^\dag\psi)^2,\\
\mathcal{O}^{(4)}&\sim(\di\psi^\dag)\psi\psi^\dag\psi-\psi^\dag(\di\psi)\psi^\dag\psi+\psi^\dag\psi(\di\psi^\dag)\psi-\psi^\dag\psi\psi^\dag(\di\psi).
\end{split}
\ee
Here $\mathcal{O}^{(1)}$ and $\mathcal{O}^{(4)}$ are odd under charge conjugation whereas $\mathcal{O}^{(2)}$ and $\mathcal{O}^{(3)}$ are even.

One subtlety at finite $N$ is that compared to large $N$ some operators are identically zero because of fermion statistics, and thus are not included in the basis. For example, for $N=3$ the quasi-primary operator $(\psi^\dag\psi)^4$ is not included in the basis. 

In order to model the multi-particle states of the single-particles, we use the bosonic quasi-primary operators
\be
\mathcal{O}_B(x^-)=\sum_{\sum {s_i}=n} b_{s_1, s_2, ..., s_{k}}\di^{s_1}(\di\phi_1)\di^{s_2}(\di\phi_2)\di^{s_3}(\di\phi_3)...\di^{s_{k}}(\di\phi_k).
\label{primaryopboson}
\ee
Each bosonic field $\phi_i$ corresponds to the excitation of a single-particle state. The $\di\phi$ operator, a singlet of $SU(N)$, is itself a quasi-primary operator of a 1+1 dimensional effective bosonic free field theory. 

The bosonic composite operator $\mathcal{O}_B$ also satisfies the Killing equation
\be
[K_-, \mathcal{O}_{B}(x^-)]=i\left((x^-)^2\partial_-+2x^-(n+k)\right)\mathcal{O}_{B}(x^-). 
\ee
If some of the single-particle-states $\di \phi_{i_1}, \di \phi_{i_2}, ..., \di \phi_{i_p}$ are identical, the operator is symmetric under the exchange of these particles and consequently the coefficients $b_{s_1, s_2, ..., s_{k}}$ are symmetric on the corresponding indices $s_{i_1}, s_{i_2}, ..., s_{i_p}$. In practice we write the bosonic operators $\di\phi$ in momentum space using mode-expansion. The Killing equation then becomes a differential equation on polynomials of the momenta $p_1, p_2, ..., p_k$ of the $\phi$ fields involving the coefficients $b_{s_1, s_2, ..., s_{k}}$. The solutions to this differential equation for the bosonic quasi-primary operators are products of Jacobi polynomials $P^{(a,1)}_m$, with $m$ and $a$ integers that label different solutions, and the arguments of the Jacobi polynomials being linear combinations of the momenta $p_i$. This is similar to the calculation of the fermionic operators and the readers are referred to \cite{Katz:2013qua} for details. Examples of the lowest dimensional bosonic operators are
\be
\begin{split}
\mathcal{O}^{(1)}_B&\sim\di\phi_1\di\phi_2,\\
\mathcal{O}^{(2)}_B&\sim\di^2\phi_1\di\phi_2-\di\phi_1\di^2\phi_2,\\
\mathcal{O}^{(3)}_B&\sim\di\phi_1\di\phi_2\di\phi_3.\\
\end{split}
\ee

At large $N$, where the interaction between the single-particle-states is suppressed, the spectrum of non-interacting multi-particles calculated using the basis of the bosonic quasi-primary operators is expected to match with the 't Hooft model multi-particle spectrum, as expected from effective conformal dominance. At finite $N$, interactions between the mesons are important and there are corrections to the energy proportional to powers of $1/N$ times the mass of the bound-states. Therefore at finite $N$ the free boson approximation to the continuum is no longer valid. Nevertheless, for multi-particle-states made of only the decoupled massless state, $|B_0\rangle$, plus a single massive single-particle meson, there are no interactions between the single-particles. This is the case described in Section \ref{multi-part}. Multi-particle-states of this kind are expected to be well identified by the free spectrum, even at finite $N$.

\section{The Density of States}
\label{app density}

\label{dos}
\begin{figure}
\begin{center}
\includegraphics[width=0.65\textwidth]{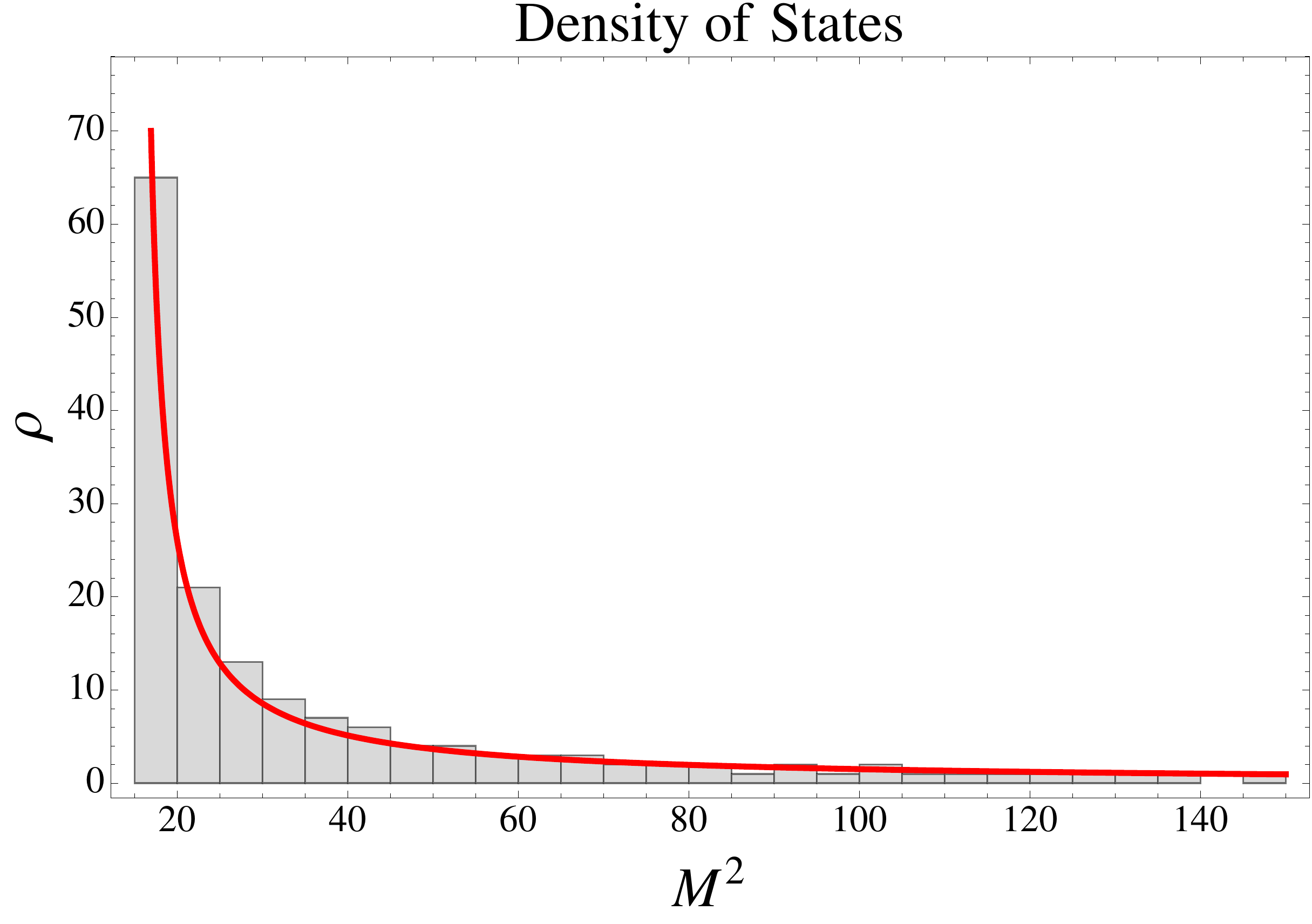}
\end{center}
\caption{Density of states of the two-body state of $|B_1\rangle$ and $|B_0\rangle$ at a high cutoff dimension $\Delta_{\text{max}}=200$. The histogram shows the counting of states with bin width $\Delta M^2=5$. It matches with the expected density of states of the two-body continuous spectrum (red line).}
\label{dosplot}
\end{figure}

The continuum spectrum can be recovered using the conformal basis approach at large cutoff dimension $\Delta_{\text{max}}$.  Because of the discreteness of the basis at a certain $\Delta_{\text{max}}$ we can only obtain a subset of the multi-particle states. But as one increases $\Delta_{\text{max}}$ the discrete states start to converge and mimic the behavior of the continuum, as is illustrated in the following example of the bosonic quasi-primary operator basis.

In Figure \ref{dosplot} we show the density of states at $\Delta_{\text{max}}=200$ for the bosonic two-body states that contain $|B_1\rangle$ and $|B_0\rangle$. The discrete spectrum is calculated using the bosonic quasi-primary operators for the non-interacting two-body mass matrix
\be
M^2(x)=\frac{M^2_{B_1}}{x},
\ee
since $|B_0\rangle$ is massless. The parton variable $x$ is integrated from 0 to 1. The counting of states, binned with respect to $M^2$, is compared with the expected density of states of the continuum. The latter is given by
\be
\rho(M^2)=\frac{Z}{M^2-M^2_{B_1}}.
\ee
Here $Z$ is a normalization determined by a fit to the distribution of counting. The fluctuation in each bin count compared to the expected density of states is within $20\%$.

\bibliography{finiteNref}{}
\bibliographystyle{JHEP}

\end{document}